# Strangers in a foreign land: 'Yeastizing' plant enzymes

Kristen Van Gelder[1], Steffen N. Lindner[2,3], Andrew D. Hanson[1], Juannan Zhou[4,*]

[1]Horticultural Sciences Department, University of Florida, Gainesville, FL 32611, USA.

[2]Department of Systems and Synthetic Metabolism, Max Planck Institute of Molecular Plant Physiology, Potsdam, Germany.

[3]Department of Biochemistry, Charité Universitätsmedizin Berlin, corporate member of Freie Universität Berlin and Humboldt-Universität, 10117 Berlin, Germany

[4]Department of Biology, University of Florida, Gainesville, FL 32611, USA.

*Corresponding author: juannanzhou@ufl.edu

# ABSTRACT

Expressing plant metabolic pathways in microbial platforms is an efficient, cost-effective solution for producing many desired plant compounds. As eukaryotic organisms, yeasts are often the preferred platform. However, expression of plant enzymes in a yeast frequently leads to failure because the enzymes are poorly adapted to the foreign yeast cellular environment. Here we first summarize current engineering approaches for optimizing performance of plant enzymes in yeast. A critical limitation of these approaches is that they are labor-intensive and must be customized for each individual enzyme, which significantly hinders the establishment of plant pathways in cellular factories. In response to this challenge, we propose the development of a cost-effective computational pipeline to redesign plant enzymes for better adaptation to the yeast cellular milieu. This proposition is underpinned by compelling evidence that plant and yeast enzymes exhibit distinct sequence features that are generalizable across enzyme families. Consequently, we introduce a data-driven machine learning framework designed to extract 'yeastizing' rules from natural protein sequence variations, which can be broadly applied to all enzymes. Additionally, we discuss the potential to integrate the machine learning model into a full design-build-test-cycle.

# Introduction

Many high-value plant compounds are end products of secondary metabolic pathways and are present in minute quantities, typically less than 1% of total dry weight (Andrea *et al.*, 2010; Paddon *et al.*, 2013; Galanie *et al.*, 2015; Pyne *et al.*, 2020; Zhang *et al.*, 2022). Thus, isolating pure compounds from plants is often not feasible as it requires large quantities of plant material and, for wild plant sources, can be detrimental to native ecosystems.

Consequently, various alternative routes for producing plant secondary compounds have been explored; these have traditionally included chemical synthesis (Toure and Hall, 2009) and plant cell or tissue culture (Atanasov *et al.*, 2015). Recently, synthetic biology approaches involving microbial cell factories have been adopted (Cravens *et al.*, 2019). Here researchers can either transplant entire pathways into microbial hosts for direct production of the final compound or integrate partial pathways into a hybrid framework that combines biological synthesis with downstream chemical synthesis (for example see Paddon *et al.*, 2013 for the synthesis of artemisinin).

As unicellular eukaryotes, the yeasts *Saccharomyces cerevisiae* (baker's yeast) and *Pichia pastoris* (also known as *Komagataella phaffii*) have cellular environments more akin to plants than bacteria. This makes them preferred platforms for establishing cell factories to produce eukaryotic proteins (Mattanovich *et al.*, 2012). Additionally, their similarity to plant cells makes yeasts suitable hosts for producing plant secondary metabolites and for characterizing these complex pathways by elucidating the reactions mediated by individual enzymes (Klonus *et al.*, 1994; Sato *et al.*, 1999; Vieira Gomes *et al.*, 2018; Boonekamp *et al.*, 2022).

The widespread use of yeasts as expression platforms for plant genes and pathways is supported by a powerful molecular genetic toolset that includes various plasmids, selection markers, promoters, genome editing tools, and tags and fusions to track protein localization.

These tools are critical for constructing and optimizing heterologous plant pathways (Jensen and Keasling, 2015), thus making it far easier to establish cellular factories in yeast than in other systems.

Typically, producing a plant secondary metabolite in yeast begins with identifying the biosynthetic pathway responsible. Once this pathway and its genes are known, the next step is to design the synthetic metabolic pathway, either by wholesale use of all the enzymes from the natural plant pathway or by constructing chimeric pathways with enzymes from different species (e.g. see production of the opioids thebaine and hydrocodone by Galanie *et al.* 2015). The pathway genes are transferred into yeast individually or as small modules, with the performance of the enzyme/pathway evaluated based on the titer, production rate, and yield of the respective pathway intermediate(s) or the final product.

Developing a yeast strain able to make the targeted product on a commercial scale can pose major challenges, particularly: (1) the functions of the enzymes and pathways may be insufficiently understood (S. Li *et al.*, 2018), leading to the construction of non-functional or suboptimal pathways; (2) the heterologous plant pathway may not integrate well into the endogenous metabolic network of the yeast host (Chen *et al.*, 2020), thereby stressing the host cells and impairing growth; (3) the heterologously expressed enzymes often have lower catalytic activity than when expressed in the native plant host (S. Li *et al.*, 2018; Chen *et al.*, 2020), resulting in metabolic bottlenecks which require substantial optimization work to overcome. The first two obstacles may be tackled using rational metabolic engineering principles. These include modifying the pathway structure by replacing subsets of enzymes or introducing additional enzymes. Additionally, host metabolic pathways may be reconfigured to increase precursor supply (e.g. Koopman *et al.*, 2012; Zhang *et al.*, 2022), or to control accumulation of toxic intermediates (Dahl *et al.*, 2013; Paddon *et al.*, 2013; Zhao *et al.*, 2018). In contrast, addressing the third obstacle often requires targeted engineering at the gene level. As enzyme inefficiency is frequently (though often without evidence) attributed to insufficient gene expression, codon optimization and promoter/gene copy number modification are usually the first recourse (Siddiqui *et al.*, 2012), with the latter often proving more successful.

Equally likely, however, is the possibility that the plant genes fail to express in yeast cells due to poor adaptation of the plant enzymes to the foreign environment in the yeast host. This type of failure will respond minimally to gene overexpression and must be salvaged through changes to the protein sequence. Potential mechanisms of enzyme failure include protein instability, improper folding, incorrect localization, insufficient substrate or cofactor concentrations, and excessively fast protein turnover (Besada-Lombana *et al.*, 2018) (Figure 1). Aside from protein localization (Kumar *et al.*, 2002; Huh *et al.*, 2003), the mechanisms determining the success or failure of functional heterologous enzyme expression in yeast are poorly understood. This knowledge gap leads to the scarcity of guiding principles for optimizing heterologous enzyme activities on the protein sequence level. To date, the only broadly applicable mechanism-agnostic strategy is directed evolution (Wang *et al.*, 2019), which is time-consuming and costly. Moreover, directed evolution results are rarely generalizable, thus necessitating a case-by-case approach. This greatly limits the rate at which high-value synthetic pathways, e.g. involving plant enzymes, can be established in yeast cell factories.

In this review, we address the conceptual and technical gaps in engineering plant enzymes for optimal performance in microbial hosts. Importantly, we argue that the plant enzymes face many common challenges when heterologously expressed in a host like yeast due to drastic differences in cellular environments (summarized in Figure 1) that lead to suboptimal enzyme activities. This implies the potential to develop generalizable solutions for modifying the protein sequence of plant enzymes for better adaptation to the yeast environment (i.e., yeastizing') that can be broadly applied to enhance performance. We further argue that despite the limited mechanistic understanding of enzyme failure, we may nonetheless extract general yeastizing rules by learning from natural sequence variation among plant and yeast proteins.

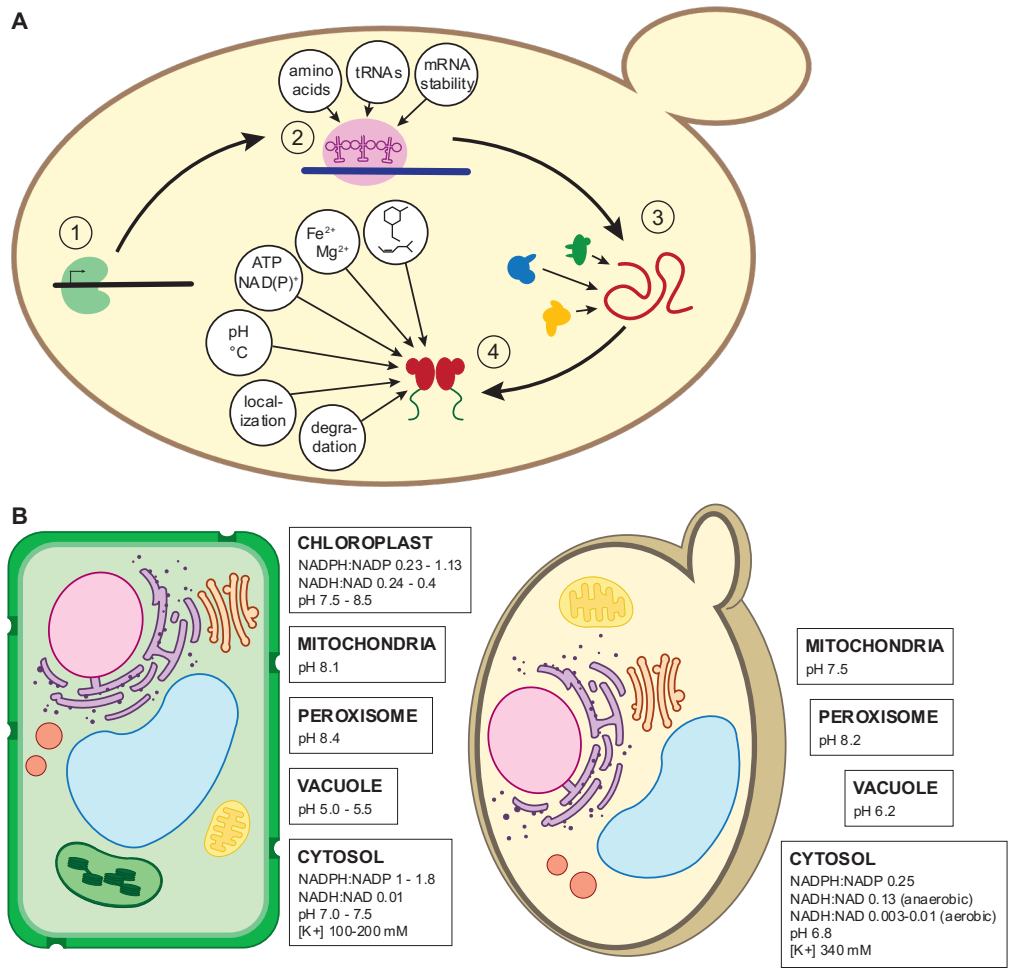

Figure 1: Cellular influences on enzyme activity which could lead to enzyme failure for plant genes expressed in yeast. **A:** Enzyme production from gene to protein is controlled by several factors that are different in plants and yeast. (1) Gene expression generates mRNA as a transcript of heterologous genes; (2) Translation depends on the availability of amino acids, tRNAs. mRNA stability effects the amount of full protein produced; (3) Protein folding and maturation is influenced by chaperones, molecular composition of the cellular environment, temperature, pH, post-translational modifications, metals, redox state, localization; (4) the function of the mature enzyme is influences by cofactor and cosubstrate availability, pH, temperature, localization and its half-life (i.e. protein degradation). **B:** comparison of environments in cellular compartments of plant and yeast cells. Data for NAD(P):NAD(P)H ratios was collected from Albe et al. (1990), Canelas et al. (2008), Van Eunen et al. (2010), Smith et al. (2021), Heineke *et al.* (1991), and Höhner *et al.* (2021). Potassium (K+) concentrations were collected from Hirsch et al. (1998), Szczerba et al. (2009), and Van Eunen et al. (2010). pH values were extracted from Bortolotti et al. (2006), Martinez-Munoz and Kane (2008) Cosse and Seidel (2021).

Below, we first summarize current approaches to enhancing plant enzyme functions in yeast and highlight their limitations. We then propose a framework for extracting yeastizing rules from natural protein sequence variations by using data-driven machine learning (ML) techniques and present a potential implementation based on the so-called generative adversarial networks (GANs). To support the feasibility of our proposed ML framework, we perform a pilot bioinformatics study that reveals systematic deviations in amino acid usage and sequence motifs between plant and yeast proteins. Finally, we discuss the possibility of integrating experimental approaches including high-throughput screening for fine-tuning the yeastizing ML model.

## Challenges in optimizing the activity of heterologously expressed plant genes and current engineering strategies

Enhancing the heterologous activity of plant proteins is one of the most prominent challenges in the development of efficient microbial cell factories. The significant disparities in the intracellular environments (as depicted in Figure 1) contribute to the common observation that naive expression of unmodified plant genes frequently results in suboptimal enzyme activities (S. Li *et al.*, 2018; Cravens *et al.*, 2019). This is supported by the many studies we cite in this section that identified rate-limiting enzymes within heterologously expressed plant pathways, which often require substantial efforts to improve their activities. Further, it should be noted that due to a survivorship bias wherein only positive results are published in peer-reviewed journals, the actual rate of failure for heterologous expression of plant genes in microbial platforms (in particular yeast) almost surely surpasses estimates derived from a systematic literature survey.

This section provides a concise overview of the common strategies for improving the activity of heterologously expressed plant enzymes. It is not intended to be exhaustive, but to. highlight the limitations of state-of-the-art methodologies and identify significant knowledge and technology gaps.

## Strategies for increasing the quantity of plant enzymes

The first strategy usually chosen to relieve rate-limiting steps in plant pathways expressed in yeast is to increase the expression level of the target protein. Codon optimization is widely used to do this. Codon usage, in particular the frequency of rare codons, impacts translational efficiency (Tuller *et al.*, 2010) and fidelity (Ikemura, 1981). Considering yeast's distinct codon usage bias (CUB) compared to plant CUBs (Gustafsson *et al.*, 2004; Hershberg and Petrov, 2008, 2009; Plotkin and Kudla, 2011; Parvathy *et al.*, 2022), a common strategy is to modify the nucleotide sequence of the plant gene to match the CUB of the yeast host. Various codon optimization methods have been developed to replace less frequently used codons in the source gene with the most commonly used codons in the host (Richardson *et al.*, 2006; Villalobos *et al.*, 2006; Feng *et al.*, 2010; Marlatt *et al.*, 2010). Due to its simplicity, codon optimization has become almost a default method when expressing plant proteins in yeast and is sometimes uniformly applied *a priori* to all heterologously expressed plant genes. For example, in an effort to synthesize the plant hormone gibberellin in the oleaginous yeast *Yarrowia lipolytica*, all enzymes from *Arabidopsis thaliana* were codon optimized (Kildegaard *et al.*, 2021). Similarly, to produce the cyanogenic glycoside dhurrin in *S. cerevisiae*, Kotopka and Smolke (2019) codon-optimized all the heterologously expressed genesIn another case, Paddon *et al.* (2013) codon-optimized all three *Artemisia annua* genes for synthesizing artemisinic acid from amorphadiene in *S. cerevisiae*,

Despite its widespread use, codon optimization is not necessarily always effective in enhancing heterologous expression of plant proteins. This may be partly due to the limitation of traditional codon optimization methods where the nucleotide sequence is recoded according to the most frequent codons in the yeast host, which may not lead to increased protein expression due to effects such as codon-mediated control of co-translational protein folding (Liu, 2020). New codon optimization methods have been designed that consider factors other than universal codon preferences (see Komar *et al.*, 1998; Morgan and Conner, 2001; Richardson *et al.*, 2006; Gaspar *et al.*, 2012; Mellitzer *et al.*, 2012; Lanza *et al.*, 2014; Fu *et al.*, 2020). For example, codon harmonization aims to mimic the natural codon usage pattern present in the source plant (Mignon *et al.*, 2018). Instead of full-force codon optimization, this strategy aims to align protein synthesis velocity to the donor strain and hence maximize correct folding, localization, and functionality of the enzyme. However, as such methods have not yet been widely applied to yeast, evidence for their superiority over traditional codon optimization is lacking.

Another common strategy to improve protein expression is to increase the copy number of the target genes. For example, to achieve high-yield production of lycopene, Ma *et al.* (2019) duplicated two key plant genes, namely geranylgeranyl diphosphate synthase (*crtE*) and phytoene desaturase (*crtI*), at the final steps of the lycopene pathway. This maneuver increased lycopene production four-fold relative to the reference strain. Likewise, for production of tropane alkaloids in *S. cerevisiae*, Srinivasan and Smolke (2020) optimized the pathway by integrating two additional copies of the putrescine N-methyltransferase 1 (*PMT*) gene from *Atropa belladonna* and *Datura stramonium*, as well as one additional copy of pyrrolidine ketide synthase (*AbPYKS*) from *A. belladonna*. This increased production of intermediates downstream of the precursor putrescine up to five-fold and tropine production over two-fold. Similarly, Rodriguez *et al.* (2017) duplicated the chalcone synthase (CHS) and chalcone reductase (*CHR*) genes in the tyrosine ammonia-lyase route of flavonoid synthesis in *S. cerevisiae* to obtain a 1.5-fold increase in the titre of fisetin.

A third common strategy to increase abundance of rate-limiting enzymes is to place the corresponding genes behind stronger promoters (Andrea *et al.*, 2010; Kildegaard *et al.*, 2021) or better ribosome binding sites (Naseri and Koffas, 2020).

Although widely used, the above strategies to tune enzyme expression levels may not be effective, due to several factors. First, protein overexpression may lead to downregulation of the target gene via negative feedback mechanisms, e.g. the classical case where overexpression of chalcone synthase in pigmented petunia petals blocks anthocyanin biosynthesis (Napoli *et al.*, 1990). Second, overexpression of the genes may impose a metabolic burden on the host cell that reduces growth rate, leading to lower product titre (Hershberg and Petrov, 2009). Third, and crucially, overproduction of the protein may still not guarantee higher product titres if the enzyme fails to carry out its native function efficiently due to a mismatch between the intracellular environment of the microbial host and the plant from which the enzyme came. Such mismatches can lead to incorrect protein folding, mislocalization, and lower catalytic efficiencies (Figure 1). In these cases, protein engineering approaches directly modifying the amino acid sequence must be employed. In the following, we review common sequence engineering strategies for rate-limiting enzymes in heterologously expressed plant pathways. We focus on approaches that aim to preserve the original function of the targeted enzyme, rather than ones that aim to further increase the enzyme's catalytic activity beyond the native level or to change the original enzyme's catalytic mechanisms, substrates, or products.

**Protein engineering strategies for preserving the native activity of heterologously expressed plant enzymes**

*Mechanism-aware approaches*

There are two basic strategies for yeastizing the primary sequence of heterologously expressed plant enzymes. The first is rational redesign of the plant sequence, based on a solid biological hypothesis about the rate-limiting enzyme. In such cases, it is often possible to introduce targeted amino acid changes to the plant enzyme to achieve higher activity.

A good example of rational redesign is the N-terminal engineering of rate-limiting enzymes. Many enzymes contain N-terminal peptides that determine the enzyme's subcellular localization (Emanuelsson *et al.*, 2000). This poses a potential challenge in ensuring the correct functioning of heterologously expressed plant genes due to either the absence of the original plant localization target (e.g. the plastid), or differences in localization signals for the same cellular compartment between plant and the new host. Subcellular protein localization can be determined through microscopy and fluorescent fusion proteins (Zhang *et al.*, 2022), or bioinformatically predicted (Emanuelsson *et al.*, 1999, 2007; Small *et al.*, 2004). Rational design principles can then be applied to excise the original target for relocalization to the yeast cytosol or to replace the original targeting peptide to direct the enzyme to a specific cellular compartment in yeast.

In the aforementioned example of gibberellin biosynthesis in yeast (Kildegaard *et al.*, 2021), the authors identified plastidial targeting signals of three genes early in the *A. thaliana* pathway converting geranylgeranyl diphosphate to ent-kaurenoic acid (copalyl diphosphate synthase, *CPS*; entkaurene synthase, *KSp*; entkaurene oxidase, *KOp*). The authors constructed truncated variants of the three genes for re-localization to the yeast cytosol. This removal of the plastid localization signals resulted in a four- to six-fold increase in the final gibberellin production. Similarly, in building a yeast cell factory for 13R-manoyl oxide, Zhang *et al.* (2019) removed N-terminal plastid targeting peptides from the *Coleus forskohlii* diterpene synthases CfTPS2 and CfTPS3. This modification, together with overexpression of the two genes using the TEF1p and TDH3p promoters, increased the production of 13R-manoyl oxide by 6.6-fold. Li *et al.* (2018) removed the 24-residue N-terminal targeting peptide from (S)-norcoclaurine synthase (NCS) using a CRISPR/CAS9 genome editing strategy. This operation led to an almost eightfold increase in noscapine production.

Other applications may require the plant protein to be localized to the equivalent yeast cellular compartment. For instance, many plant cytochrome P450 enzymes have N-terminal signal peptides that correctly anchor the nascent polypeptide chain to the endoplasmic-reticulum (ER) membrane (Gnanasekaran *et al.*, 2015). Consequently, researchers have sought to improve plant P450 activities in yeast by modifying the native N-terminal sequence. For example, to achieve opioid biosynthesis in yeast, Galanie *et al.* (2015) constructed chimeric proteins for a key pathway P450 (salutaridine synthase) by attaching the N-terminal α-helices from a similar enzyme cheilanthifoline synthase (CFS) isolated from two *Papaver* species. This resulted in in a six-fold increased conversion rate of (R)-reticuline to salutaridine.

*Localization to new compartment*

Researchers have also attempted to direct an enzyme to non-cytosolic locations by adding novel targeting peptides. For instance, Lyu *et al.* (2019) attempted to relocalize the *Citrus* flavanone 3-hydroxylase (F3H) and *Arabidopsis* flavanol synthase (FLS) enzymes to yeast mitochondria to increase access to the co-substrate 2-oxoglutarate. To achieve this, a 26-residue N-terminal mitochondrial localization signal from yeast cytochrome oxidase (CoxIV) was fused to both F3H and FLS. However, this unexpectedly caused a sharp decrease in kaempferol production.

*Mechanism-agnostic approaches*

The factors compromising the performance of rate-limiting enzymes in heterologous hosts are sometimes unknown. In these cases, researchers often resort to screening candidate gene variants or to protein evolution to increase activity. Although these approaches have the advantage of not requiring mechanistic understanding about the targeted enzyme they have some major limitations. First, the lack of guiding rational design principles often requires screening mutations across the entire protein, instead of a subset of high priority positions, which greatly increases the screening effort needed. Second, successful results are generally not generalizable to other enzymes. Thus, screening must be performed for all rate-limiting enzymes in each pathway on a case-by-case basis. To conclude this section, we provide an overview of the common mechanism-agnostic strategies for protein engineering to enhance enzymatic activities.

Screening natural protein variants is among the earliest and commonest procedures for optimizing enzyme activities. For example, in an effort to establish anthocyanin production in *S. cerevisiae*, Eichenberger *et al.* (2018) screened three to nine natural plant variants of flavanone-3-hydroxylase (F3H), flavonoid-3′-hydroxylase (F3′H), dihydroflavonol-4-reductase (DFR), and anthocyanidin-3-O-glycosyl transferase (A3GTs). Similarly, Lyu *et al.* (2019) screened four plant variants for F3H and FLS to identify the optimal combination for production of kaempferol in *S. cerevisiae*. Zhang *et al.* (2022) improved the activity of the rate limiting Pictet–Spengler-type reaction catalyzed by the enzyme strictosidine synthase (STR) by screening seven STR homologs identified in monoterpene indole alkaloid (MIA)-producing plants; the best variant tested had more than ten-fold higher activity than the poorest. In a study by Luo *et al.* (2019), which achieved complete biosynthesis of cannabinoids in yeast, six *Cannabis* and two *Humulus lupulus* prenyltransferases were screened for their activity in catalyzing the alkylation of geranyl pyrophosphate (GPP) and olivetolic acid (OA) to form the cannabinoid precursor cannabigerolic acid (CBGA). Interestingly, the previously patented enzyme CsPT1 was shown to produce CBGA in insect cells (Page, 2012) but not when expressed in yeast, indicating the possible role of intracellular environments in modulating enzyme activities.

There are two limitations with screening natural plant variants for enzymatic activities in heterologous hosts. First, compared with bacterial and mammalian genomes, plant genomes are poorly annotated (Kersey, 2019), which can limit the number of candidate variants available for testing. Here, AlphaFold can assist to improve poorly annotated genomes by predicting a 3D structure of the protein (Jumper *et al.*, 2021). This structure can be a query to screen for structurally similar proteins, e.g. by using Foldseek (van Kempen *et al.*, 2023).

Second, while in the examples above the plant homologs exhibited a fairly wide range of activities when expressed in yeast, all plant natural variants are well-adapted to plant cellular environments, which are much more similar to each other than to the environment in a yeast cell (Figure 1). Thus, natural variants of plant enzymes may show too little variation to allow researchers to find candidates with adequate preadaptation to the yeast cellular environment. This would apply particularly to plant enzymes that depend absolutely on a cellular factor such as a specific cofactor or post-translational modification, leading to complete lack of activity and zero variation among homologs when expressed in yeast.

An orthogonal strategy to overcome this challenge is to construct and test completely synthetic sequences through screening or directed evolution experiments (Cobb *et al.*, 2013; McLure *et al.*, 2022). In an early study utilizing high-throughput assays, Alberstein *et al.* (2012) developed a color-based screen for directed evolution of phenylpropanoid pathway enzymes. The assay is based on the read-out of the colored intermediate naringenin that is synthesized from the end product, 4-coumaroyl-CoA, of the phenylpropanoid pathway and malonyl-CoA by chalcone synthase (CHS). The authors screened ~50,000 4-coumarate:CoA ligase (C4L) variants generated by PCR mutagenesis. The highest-activity variant increased product yield about four-fold compared with the wild-type enzyme. In another example, DeLoache *et al.* (2015) established production in yeast for the key benzylisoquinoline alkaloids (BIA) intermediate (S)-reticuline through the intermediate L-DOPA. The authors focused on enhancing L-tyrosine hydroxylation to L-DOPA by the P450 enzyme 76AD1 (CYP47AD1) through screening ~200,000 variants generated by PCR mutagenesis. The high-throughput screening was enabled by a fluorescence assay that converts L-DOPA reaction product to the highly fluorescent pigment betaxanthin. Two major activity enhancing mutations were identified, which when combined improved L-DOPA yields by 2.8-fold compared with the wild-type CYP76AD1. In an effort to improve linalool production in *S. cerevisiae,* Zhou *et al.* (2020) took advantage of the competition between monoterpenes and carotenoids for the common precursor GPP, and developed a colorimetric assay using a yeast strain expressing the lycopene pathway. Negative correlation between lycopene level and linalool synthase activity allowed the authors to perform directed evolution of linalool synthase to obtain the variant t67OMcLIS, that, combined with overproduction of GPP by the upstream mevalonate (MVA) pathway, increased product titre more than two-fold.

High-throughput screening assays are a very promising approach for optimizing heterologously expressed plant enzymes. Furthermore, computational methods, such as supervised ML (e.g. Zhou and McCandlish, 2020; Luo *et al.*, 2021; Tareen *et al.*, 2022; Zhou *et al.*, 2022) may complement experimental results by modeling the activity landscape of all protein sequences to facilitate the identification of novel variants in the test pool. An alternative approach for optimizing enzymes for a different host organism involves the utilization of ancestral sequence reconstruction. By reconstructing the evolutionary history of enzymes, this method develops biocatalysts with enhanced properties, such as higher stability, efficiency, and adaptability to new environments. This method improves the enzymes, offering a more ideal foundation for high-throughput screening techniques (Furukawa *et al.*, 2020; Pinto *et al.*, 2022). Despite its potential, a major bottleneck in designing high-throughput assays is that an easy activity read-out such as growth or color is often not readily available for non-essential genes, thus assays must be developed on an enzyme-by-enzyme basis. Recent advances in biosensor development may streamline the engineering of enzyme-specific read-outs (Sarnaik *et al.*, 2020). However, major challenges remain to increase the sensitivity and dynamic range of current biosensors to render this approach widely applicable. Alternatively, activity of the target enzyme can be coupled to growth of the host to allow efficient selection of beneficial mutants through competition among mutant strains. This strategy has been successfully applied to bacteria, wherein a strain can be engineered to depend on a key molecule through gene deletion and selected for metabolic flux through the target enzyme or pathway (Zhang *et al.*, 2018; Kramer *et al.*, 2020; Orsi *et al.*, 2021). However, implementing this so-called growth-coupled selection strategy for plant secondary metabolic enzymes may not be straightforward due to the non-essentiality of the focal plant compounds to the fungal host. Finally, recent technological advances in continuous directed evolution in yeast (Ravikumar *et al.*, 2018; García-García *et al.*, 2022; Molina *et al.*, 2022) combine somatic hypermutation and uninterrupted enzyme evolution. Such methods can also be integrated with high-throughput assays to ensure faster evolution of the target enzyme towards higher activity through direct competition between mutant strains, surface display (Wellner *et al.*, 2021), or fluorescence-assisted cell-sorting (Javanpour and Liu, 2021).

## Potential of machine learning for yeastizing plant enzymes

In this section, we propose a general-purpose framework for yeastizing the amino acid sequence of rate-limiting enzymes in heterologously expressed plant pathways. Here it is helpful to distinguish two levels of factors determining the proper functioning of foreign enzymes in new hosts. The first are universal factors related to adaptation of an enzyme to the broad cellular environments, including mechanisms shown in Figure 1. In addition to these universal factors, there may also be cellular factors affecting enzyme function specific to the gene of interest, including presence of cofactors, ligands, interacting partners, etc. In this review, we hypothesize that (1) modifying the plant amino acid sequence for better adaptation to the first class of factors universal to all enzymes may be sufficient to allow successful expression of the plant gene in yeast; (2) divergent sequence features between the two clades largely reflects adaptation of different enzyme families to these shared cellular environmental factors, which may be learned through data-driven methods to extract yeastizing rules to facilitate successful expression of plant secondary metabolic enzymes. Based on these two hypotheses, we propose a yeastizing framework utilizing recent advances in unsupervised ML methods. The proposed method in its most basic form only relies on natural amino acid sequence variations between plant and fungi for learning adaptive sequence features but can be enhanced by integrating high-throughput experimental screening data and developed into a full design-build-test-learn cycle.

This framework is intrinsically mechanism-agnostic. We argue that given the complexity of cellular environments and the current lack of thorough understanding of biological factors affecting the success of heterologous expression of foreign enzymes, no mechanism-driven model can yet account for all the potential factors and their interactions, such that phenomenological black box ML models are the better option for learning complex yeastizing rules.

**Plant and fungal enzymes show distinct sequence features**

Before designing a yeastizing ML model relying on natural amino acid sequences from the two clades, we must first show that natural plant and yeast sequences indeed have features distinct enough to serve as the basis for yeastizing rules. To do this, we first curated a dataset consisting of amino acid sequences of 17 diverse cytosolic primary metabolic enzymes (Supplemental Table 1) with homologs in both plants and fungi. For each enzyme, we prepared sequences for 99 representative angiosperm species, and 21 fungal species including *S. cerevisiae* and its close relatives, as well as a few distantly related fungal species with high economic value (Supplemental Table 2). The sequences for the plant species were derived from the 1KP project (One thousand plant transcriptomes and the phylogenomics of green plants, 2019), whereas the sequences for the individual fungal species were obtained by blastp searches at NCBI.

To test if the two clades show distinct amino acid sequence features, we fitted three models with increasing complexity to the data and assessed the model's ability to predict the clade of origin (plant vs. fungal) for the held-out test sequences. A high prediction accuracy indicates that there are generalizable sequence features distinguishing plant vs. yeast enzymes, which can potentially be exploited to yeastize plant enzymes. The first two methods are simple logistic regression models, where the inputs are prepared by converting the raw amino acid sequences into sequence features and the probability of a sequence being of plant origin is modeled as a logistic function of a weighted sum of the input features. The first logistic regression model only relies on amino acid usage information and is trained on amino acid count data converted from the raw amino acid sequences. The second logistic regression model converts the sequence to counts of all 400 consecutive amino acid dimers and uses the dimer counts for prediction. In addition to the simple logistic regression, we also fitted a convolutional neural network (CNN), which is commonly used for modeling image or text data. Unlike the first two methods that rely on simple sequence features, the CNN performs convolutional operation in several layers to distill covariation in amino acid residues on the local as well as global level. It thus allows us to examine if there are distinctive higher-order sequence features differentiating the two clades. Specifically, the CNN model consists of four 1-dimensional convolutional layers, with kernel size = 5, and 4 channels. A dropout rate = 0.25 was applied to avoid model overfitting. Binary cross entropy loss was used to train the model at learning rate = 0.001, using the Adam optimizer. The logistic regression models were trained in the Python package scikit-learn 1.3.0. The CNN model was trained in PyTorch 1.12. We assess the model performance on the test set using the AUROC score (area under the receiver operating characteristic curve), with AUROC > 0.5 indicating better than random prediction.

We employed a rigorous leave-one-out cross-validation approach to assess the generalizability of sequence features across enzyme families. In each iteration, we partitioned the entire dataset into a training set, encompassing all plant and fungal sequences for 16 out of the 17 enzymes, and a test set, comprising all sequences belonging to the one enzyme that was withheld. Since the model had not been exposed to any sequences from the test enzyme during training, it must rely on features acquired from the other enzymes in the training set to make predictions. The ability to achieve predictions superior to random chance thus supports the hypothesis that there exist common sequence features shared among enzymes, which may be learned for yeastization tasks.

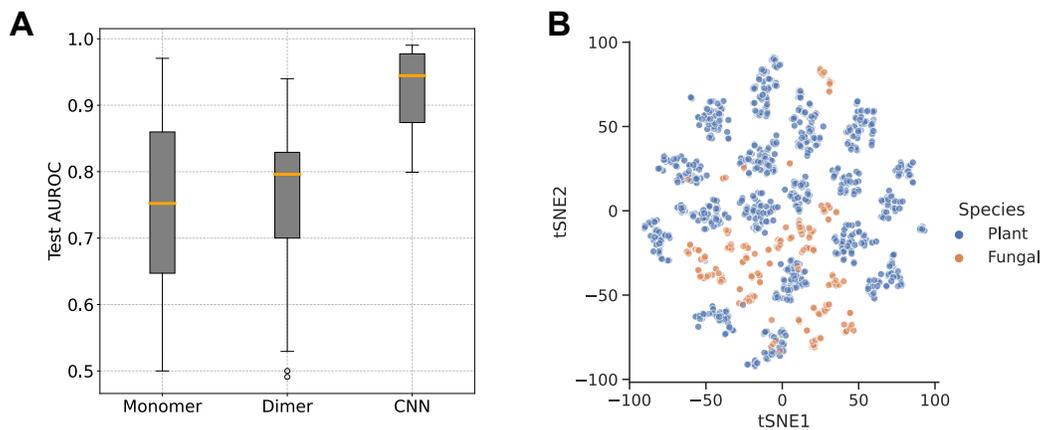

Figure 2. Natural yeast and plant enzymes show distinct amino acid sequence features. A: performance of three models at classifying the clade of origin (plant vs. fungal) of protein sequences of 17 enzyme families. The models include logistic regression using amino acid monomer and dimer counts, and a convolutional neural network (CNN). Predictive accuracy is summarized using out-of-sample AUROC for the focal held-out enzyme family by model trained on the rest of 16 families. B: Low dimensional t-SNE embeddings of the amino acid dimer count data showing sequences for the 17 enzyme families and two clades (plant vs. fungal) form distinct clusters. Dots represent protein sequences.

In Figure 2A, we present the leave-one-out AUROC scores for all three models. Notably, even the basic amino acid usage model achieved an average test AUROC of 0.75. With the more complex dimer count model, we were able to raise the mean AUROC to 0.8. Finally, using the more sophisticated CNN model pushed our mean AUROC score to 0.96. To understand why our models perform well on distinguishing plant vs. fungal sequences, we performed dimensionality reduction of the sequence features for all protein sequences. Briefly, we used t-SNE (der Maaten and Hinton, 2008) to reduce the 400 consecutive dimer count features to two dimensions. In Figure 2B, we first see that different enzyme families form distinct clusters in this low dimensional representation. Importantly, within each enzyme family, plant and fungal sequences form largely nonoverlapping groups, which makes it possible to accurately assign clade labels using straightforward models like logistic regression.

These results provide strong pilot evidence for the presence of distinct sequence features in fungal and plant proteins shared among enzyme families. Furthermore, our findings suggest that while some sequence features (such as amino acid usage and dimer counts) are intuitively understandable, higher-level features involving complex covariation among residues extracted by more sophisticated models such as CNN are also critical for differentiating the two clades.

**Proposed machine learning framework for extracting yeastizing rules from natural sequences**

The results in the previous section show that plant and fungal enzymes have distinct sequence features that are highly generalizable across enzyme families. Conceptually, it may be helpful to envision plant and fungal enzymes as inhabiting distinct "regions" within the broader protein feature space (illustrated in Figure 3A), such that homologs from these two clades across different enzyme families differ systematically along certain general directions. This suggests the possibility of mining the natural plant and yeast sequences to extract this information in the form of yeastizing rules that can be applied to secondary plant metabolic enzymes to favor better expression in yeast.

Recently, ML methods have been remarkably successful in the analysis of biological sequences in many fields. In this review, we highlight the potential of using ML to develop computational yeastizing pipelines based on natural plant and fungal protein sequences. This approach can be naturally integrated into the broader effort of heterologous expression of plant pathways through cost-efficient *in silico* redesign of rate-limiting enzymes (Figure 3B).

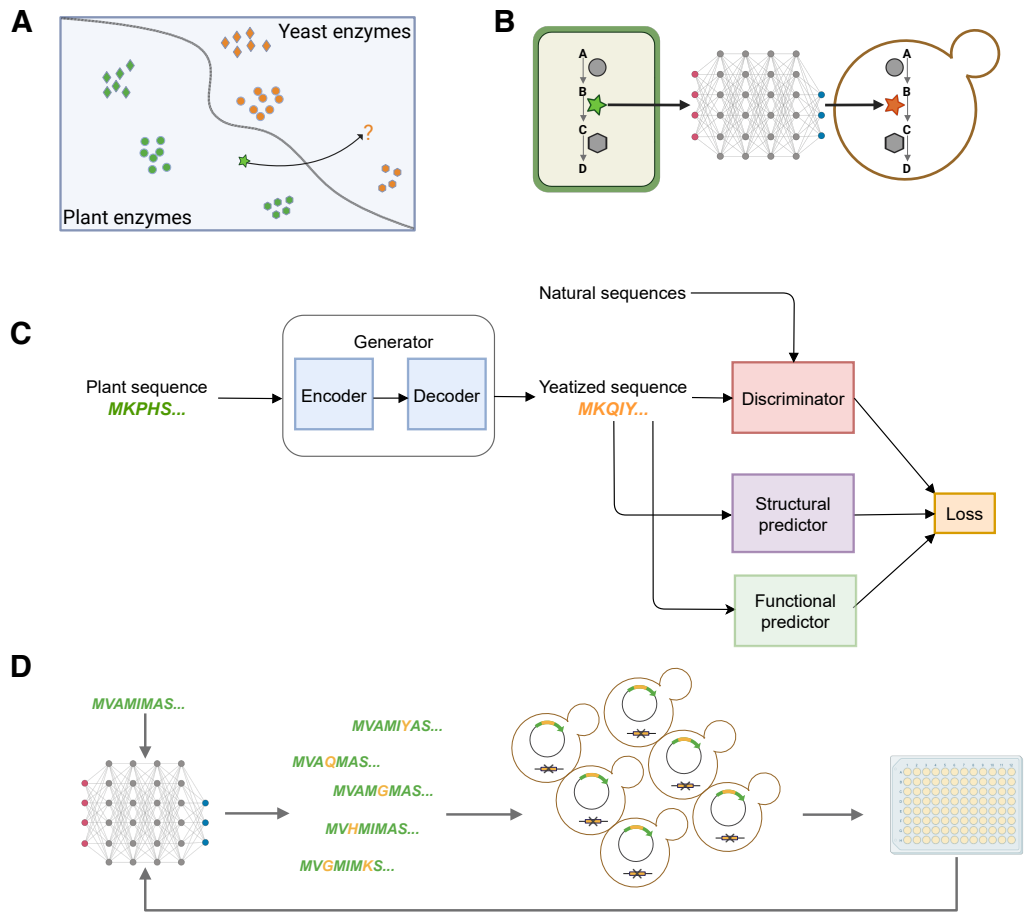

Figure 3. Machine learning framework for extracting yeastizing rules from natural sequences. (A) Hypothetical drawing showing natural plant and yeast enzymes with systematic, distinct sequence features across different enzyme families, which may be leveraged for yeastizing a plant enzyme without yeast homologs (star). (B) Machine learning model implementing the yeastizing rules may be used to optimize rate-limiting enzymes in heterologously expressed plant pathways. (C) An example of a generative adversarial network architecture for learning the yeastizing rules from natural plant and yeast sequences. (D) A design-build-test-learn cycle for incorporating functional constraints into the unsupervised yeastizing model. A plant sequence is input to the yeastizing model to propose variants carrying different mutations. The variants are transformed to yeast KO strains. Performance of different yeastized variants is measured using growth assays. The results are used to fine-tune the trained model.

Building on our hypothesis that the divergence in sequence features between the two clades largely corresponds to protein adaptation to distinct cellular environments in plants and fungi, we next propose an unsupervised ML framework to utilize natural plant and fungal sequences for yeastizing plant enzymes. While there are many unsupervised neural network architectures that are potentially suitable for yeastizing plant protein sequences after training on natural sequences, here we give an example of a simple generative adversarial neural network (GAN). A GAN is a ML architecture comprising a generator and a discriminator. The generator creates synthetic data samples, and the discriminator evaluates their authenticity. Through iterative competition, the generator is trained to generate realistic samples that deceive the discriminator. Consequently, the discriminator improves its ability to differentiate between real and fake samples. This adversarial process drives the GAN to generate high-quality synthetic data that closely resembles features of the real data. One notable advantage of the GAN architecture is its independence from paired inputs, as required in sequence-to-sequence (seq2seq) translation tasks. This flexibility allows us to expand the training dataset by incorporating plant proteins lacking yeast homologs.

To train a GAN model capable of translating a plant protein sequence to a yeastized version, we built our model on the conditional GAN architecture (Mirza and Osindero, 2014) with the architecture shown in Figure 3C. Specifically, a plant sequence is passed into the model and converted to a synthetic sequence using existing encoder-decoder architecture such as transformers. The synthetic sequence and a random real plant or yeast sequence are then simultaneously passed to the discriminator to predict the true class labels (synthetic vs. true plant vs. true yeast). Through iterative training using natural plant and yeast protein sequences, the model will learn to modify plant sequences to closely mimic natural yeast enzymes, thus achieving the yeastizing task.

A limitation of this unsupervised ML architecture is that the model is solely trained to make a plant sequence superficially yeast-like. To address this limitation, we emphasize two complementary approaches to ensure the structural and functional integrity of the yeastized sequences that can be naturally incorporated into the ML framework above.

The first strategy is designed to ensure the predicted protein folds to the native 3D structure. This step involves passing the predicted sequence from the GAN model to a structural prediction ML network such as AlphaFold2 (Jumper *et al.*, 2021). The resulting predicted structure is then compared with the ground-truth 3D structure of the input sequence to evaluate any loss of structural integrity, which can be incorporated into our overall loss calculation (Figure 3C). By incorporating this procedure, we can yeastize plant sequences while maintaining the original protein's 3D structure.

The second strategy aims to preserve the enzyme's function. The plant enzyme to be yeastized may contain various residues essential for catalysis, regulation, and protein-protein interaction. Modification of these residues will likely lead to enzyme failure. Thus, functional constraints may be introduced such that the ML model is forbidden to change these residues. Incorporating this constraint to the training procedure may involve overriding the model predicted residues with the fixed residues in the GAN generator or introducing additional strong loss terms to penalize changes at the critical positions. Several methods can be employed to compile the list of forbidden residues. First, a thorough literature review can provide sufficient information on key residues for well-studied enzymes. Alternatively, computational methods can be used to predict functionally important residues. For example, key residues (e.g. active site residues) can be identified by calculating the conservation score using a multiple sequence alignment containing the focal enzyme's homologs (Waterhouse *et al.*, 2009). Additionally, molecular dynamic/docking simulation and recent ML models can be used to identify active/binding sites (Singh *et al.*, 2011). It is important to bear in mind that inaccuracies in the predicted importance for residues in the forbidden list may restrict the model's ability to efficiently yeastize an enzyme. Thus, in practice, a soft constraint (e.g. allowing the model to make a minimum number of mismatches) may be preferable.

A notable strength of the proposed ML framework is that it only relies on readily available natural protein sequences found in databases like UniProt. However, two limitations of this framework should be acknowledged. Firstly, natural sequences have a mixture of contingent phylogenetic sequence signals and truly adaptive sequence features. The confounding phylogenetic factor can lead to generation of spurious amino acid substitutions by our model that may lessen its ability to effectively yeastize plant enzymes. Secondly, although we can use the above strategies to introduce functional constraints to our ML model, the effectiveness of this procedure may be limited since literature review or computational methods is unlikely to identify all residues essential for enzyme function. To conclude this section, we underscore the potential of integrating the unsupervised ML framework with experimental methods into a full design-build-test-learn (DBTL) cycle to enhance the performance of the pretrained yeastizing models. Considering the expensive and labor-intensive nature of directly assessing the function of secondary metabolic enzymes through techniques like mass spectrometry, we propose a surrogate approach relying on primary metabolic enzymes in conjunction with high-throughput growth assays for fine-tuning the unsupervised ML model.

Our proposed strategy begins by selecting a diverse set of essential primary metabolic enzymes from plants and engineering the corresponding yeast knockout strains (Figure 3D). A library of high-likelihood yeastized variants can be selected based on model prediction, synthesized, and transformed into the corresponding yeast knockout strains. The complementing ability of these yeastized plant enzymes can be conveniently evaluated through cost-effective growth assays. The performance of different yeastized enzymes provides critical information on enzyme functions of the generated sequences and could be used to refine the ML model (Figure 2D). One technical challenge in implementing this strategy is that the proposed ML model is unsupervised by design, thus cannot readily take labeled data generated by the growth assays. A possible solution is to train a supervised protein function prediction module based on the new data and use it in conjunction with the structural prediction module to ensure the structural and functional integrity of the generated sequence (Figure 3C). While there are many possible solutions for building a protein function prediction model, here we highlight the potential of exploiting recent advances in pretrained protein language models (LMs). Protein LMs such as ESM (Rives *et al.*, 2021) were trained on a large corpus of natural protein sequences, thus carry rich evolutionary, structural, and functional information in the model's latent embeddings. Recent applications show that accurate prediction of mutant effects can be achieved by training a supervised model with the pretrained embeddings as features using just a handful of labeled sequences (i.e., few-shots prediction) (Meier *et al.*, 2021). Thus, using this framework may allow us to better utilize the limited number of variants assessed in our growth assays. Further, it may allow the model fine-tuned on primary metabolic enzymes in the DBTL cycle to be more suitable for yeastizing plant secondary metabolic enzymes in our final application.

## Concluding remarks

Heterologous expression of plant secondary metabolic enzymes in yeasts and other microbial platforms has received much research effort in the last few decades. However, a major roadblock in successful construction of yeast cell factories for efficient production of high-value plant compounds is the prevalence of rate-limiting enzymes. In this review, we first outlined current approaches for optimizing rate-limiting enzymes in heterologously expressed pathways. To date, the most prevalent strategies have focused on increasing the expression level of limiting enzymes by codon optimization or gene overexpression. However, an equally likely cause of failure is that the enzyme fails to carry out its intended function due to poor adaptation to the cellular environment of the new host. While there are many factors that could contribute to enzyme failure, due to the lack of thorough understanding of these mechanisms, current protein engineering approaches (except for protein N-terminal engineering) have been mechanism-agnostic and focused on variant screening or directed evolution. Besides being costly and time-consuming, these methods yield few if any generalizable findings, so that researchers must work from scratch on an enzyme-by-enzyme basis.

To address this limitation, we propose the application of unsupervised ML methods for the *in silico* redesign (yeastization) of plant enzymes. Our proposed framework is based on the hypothesis that plant and fungal protein sequences show systematic differences in sequence features across enzyme families, which to a large extent reflect adaptations to the distinct cellular environments of the two clades. We provide evidence for the first part of this hypothesis by showing that simple ML models trained to distinguish plant vs. fungal enzyme sequences can be accurately generalized for classifying held-out test enzyme families. Based on this finding, we proposed a ML framework using the conditional GAN architecture. The proposed model can be trained solely on natural plant and fungal enzyme sequences and is designed to produce yeastized sequences that closely mimic natural fungal enzymes. We further highlight the benefit of incorporating a protein structure prediction module and the potential of developing a full DBTL cycle to iteratively fine-tune the yeastizing model using high-throughput experimental growth assays to ensure the functional integrity of the yeastized enzymes. Last, although our discussion has focused on adapting plant enzymes for expression in yeast, the same approach may be broadly applied to gene/pathway transplantation between any two clades (e.g., plantizing bacterial genes).

# Acknowledgements

We thank Pablo Nikel, Markus Ralser, Vincent Martin, Aymerick Eudes, Hector Garcia Martin, Sakkie Pretorius, Guillaume Beaudoin, Bingyin Peng and Claudia Vickers for inputs on the manuscript. The work of A.D.H. and K.V.G. was supported primarily by the U.S. Department of Energy, Office of Science, Basic Energy Sciences under Award DE-SC0020153, and by USDA NIFA Hatch proj-ect FLA-HOS-005796 and an Endowment from the C.V. Griffin, Sr. Foundation. The work of J.Z. was supported by the University of Florida College of Liberal Arts and Sciences.